\documentclass[12pt]{article}
\usepackage{amssymb}
\usepackage{amsmath}
\usepackage{amstext}
\usepackage{graphicx,epsfig}
\usepackage{epsfig}
\usepackage{verbatim} 
\usepackage{fancybox}
\usepackage{color}
\usepackage{ulem}
\usepackage{enumitem}
\usepackage{subfigure}
\usepackage{bbm}
\usepackage{parskip}
\usepackage{dsfont}
\usepackage[numbers,sort&compress]{natbib}
\usepackage{cancel}
\usepackage{verbatim}

\usepackage{verbatim}

\usepackage{fancyhdr}	
\usepackage{indentfirst}	
\usepackage{amsmath,amssymb,latexsym,epsf,epsfig,graphicx,tikz}
\usepackage[utf8]{inputenc} 
\usepackage{empheq}		
\usepackage{amsthm}
\usepackage{amsmath}
\usepackage{eucal}
\usepackage{mathtools}
\usepackage{mathrsfs}
\usepackage{yfonts}
\usepackage{array}		

\usepackage{tikz}
\usetikzlibrary{arrows,shapes}
\usetikzlibrary{trees}
\usetikzlibrary{matrix,arrows} 				
\usetikzlibrary{positioning}				
\usetikzlibrary{calc,through}				
\usetikzlibrary{decorations.pathreplacing}  
\usepackage{pgffor}							

\usetikzlibrary{decorations.pathmorphing}

\newcommand{\D}{{\rm d}}

\newcommand{\upleft}[2]{\prescript{(#1)}{}{\! #2}} 

\newcommand{\Comment}[1]{{}}
\definecolor{darkblue}{rgb}{0.95,0.05,0.05}
\definecolor{reddish}{rgb}{0.95, 0.05, 0.05}
\usepackage[linktocpage=true]{hyperref}
\hypersetup{
colorlinks=true,
citecolor=darkblue,
linkcolor=reddish,
urlcolor=darkblue,
pdfauthor={},
pdftitle={},
pdfsubject={}
}

\def\({\left(}
\def\){\right)}

\newcommand{\beq}{\begin{equation}}
\newcommand{\eeq}{\end{equation}}
\newcommand{\be}{\begin{equation}}
\newcommand{\ee}{\end{equation}}
\newcommand{\bea}{\begin{align}}
\newcommand{\eea}{\end{align}}

\newcommand{\rd}{{\rm d}}

\def\gsim{ \lower .75ex \hbox{$\sim$} \llap{\raise .27ex \hbox{$>$}} }
\def\lsim{ \lower .75ex \hbox{$\sim$} \llap{\raise .27ex \hbox{$<$}} }

\def\beq{\begin{eqnarray}}
\def\eeq{\end{eqnarray}}

\def\mpl{M_{\rm Pl}}

\def\p{{\cal P}}

\def\L*{{\cal L}_*}
\def\L{\mathcal{L}}
\def\({\left(}
\def\){\right)}

\def\p{\partial}

\def\p{\partial}

\def\<{\langle}
\def\>{\rangle}

\def\xyma{\xymatrix@M.7em}
\def\xymas{\xymatrix@M.1em}
\newcommand{\ba}{\begin{eqnarray}}
\newcommand{\ea}{\end{eqnarray}}

\usepackage{tikz}
\usetikzlibrary{decorations}
\pgfdeclaredecoration{complete sines}{initial}
{
    \state{initial}[
        width=+0pt,
        next state=upsine,
        persistent precomputation={\pgfmathsetmacro\matchinglength{
            \pgfdecoratedinputsegmentlength / int(\pgfdecoratedinputsegmentlength/\pgfdecorationsegmentlength)}
            \setlength{\pgfdecorationsegmentlength}{\matchinglength pt}
        }] {}
    \state{upsine}[width=\pgfdecorationsegmentlength,next state=downsine]{
        \pgfpathsine{\pgfpoint{0.25\pgfdecorationsegmentlength}{0.5\pgfdecorationsegmentamplitude}}
        \pgfpathcosine{\pgfpoint{0.25\pgfdecorationsegmentlength}{-0.5\pgfdecorationsegmentamplitude}}
    }
    \state{downsine}[width=\pgfdecorationsegmentlength,next state=upsine]{
        \pgfpathsine{\pgfpoint{0.25\pgfdecorationsegmentlength}{-0.5\pgfdecorationsegmentamplitude}}
        \pgfpathcosine{\pgfpoint{0.25\pgfdecorationsegmentlength}{0.5\pgfdecorationsegmentamplitude}}
}
    \state{final}{}
}

\title{}
\author{}

\numberwithin{equation}{section}

\begin{document}
%
\renewcommand{\thefootnote}{\fnsymbol{footnote}}
~
%
%
\begin{center}
{\Large \bf{Stability of Geodesically Complete Cosmologies}} 
\end{center} 
 \vspace{1truecm}
\thispagestyle{empty} 
\centerline{
{\large {Paolo Creminelli,${}^{\rm a}$}}
{\large {David Pirtskhalava,${}^{\rm b}$}}
{\large {Luca Santoni${}^{\rm c}$ and}}
{\large {Enrico Trincherini${}^{\rm c}$}}
}

\vspace{.5cm}

\centerline{{${}^{\rm a}$\it Abdus Salam International Centre for Theoretical Physics (ICTP)}}
\centerline{{\it Strada Costiera 11, 34151, Trieste, Italy}}
\vskip 0.1cm
\centerline{{${}^{\rm b}$\it Institute of Physics, \'Ecole Polytechnique F\'ed\'erale de Lausanne}}
\centerline{{\it CH-1015, Lausanne, Switzerland}}
\vskip 0.1cm
\centerline{{${}^{\rm c}$\it Scuola Normale Superiore, Piazza dei Cavalieri 7, 56126, Pisa, Italy}}
\centerline{{\it INFN -- Sezione di Pisa, 56200, Pisa, Italy}}

 \vspace{.8cm}
\begin{abstract}
\noindent

We study the stability of spatially flat FRW solutions which are geodesically complete, i.e.~for which one can follow null (graviton) geodesics both in the past and in the future without ever encountering singularities. This is the case of NEC-violating cosmologies such as smooth bounces or solutions which approach Minkowski in the past. 
We study the EFT of linear perturbations around a solution of this kind, including the possibility of multiple fields and fluids. One generally faces a gradient instability which can be avoided only if the operator $\upleft{3}{R} \delta N$ is present and its coefficient changes sign along the evolution. This operator (typical of beyond-Horndeski theories) does not lead to extra degrees of freedom, but cannot arise starting from any theory with second-order equations of motion. The change of sign of this operator prevents to set it to zero  with a generalised disformal transformation.

\end{abstract}

\newpage

\setcounter{tocdepth}{2}
\newpage
\renewcommand*{\thefootnote}{\arabic{footnote}}
\setcounter{footnote}{0}

\section{Introduction}

Inflation, the simplest and most compelling description of the early universe, is in general past-incomplete \cite{Borde:2001nh}. To describe what happens before inflation one has to face singularities, which are sensitive to the UV completion of General Relativity. However, not much progress has been made so far in resolving this kind of singularities in string theory. One way to avoid the problem is to qualitatively modify the early universe in such a way that going to early times ceases to correspond to going to higher and higher energy. If the Null Energy Condition (NEC) is satisfied, the Hubble parameter $H(t)$ is a decreasing function of time in an expanding universe and becomes singular when going to the asymptotic past. On the other hand, if the NEC is violated one can consider solutions which  always remain regular, sidestepping the quantum gravity regime. One can surely question this ``ostrich approach": why should the universe  care for our---hopefully temporary---ignorance about the UV completion of General Relativity? It is however interesting to study these alternative scenarios for the very early universe for various reasons. For instance, NEC violation seems mandatory to dynamically relax the cosmological constant \cite{Abbott:1984qf,Steinhardt:2006bf,Alberte:2016izw}. 

\vskip 0.15cm

Models which violate the NEC are usually characterised by instabilities, but in recent years a few counter-examples to this common lore were found (see for instance \cite{Creminelli:2006xe,Nicolis:2009qm}): leaving aside the issue of UV completion, one cannot find any general pathology associated to NEC-violation. 
As a further step one may wonder whether some kind of pathology appears when considering models with a transition between NEC-violating and standard, NEC-preserving, phases \cite{Elder:2013gya,Libanov:2016kfc, Kobayashi:2016xpl,Ijjas:2016tpn,Kolevatov:2016ppi,Ijjas:2016vtq}.
 Indeed, in many examples one is able to push the instabilities outside the NEC-violating region, but they still show up somewhere once the whole history of the universe is considered. 
 
\vskip 0.15cm 
 
 In this paper we come back to this problem using an EFT approach. Namely, we will study (Section \ref{formalism}) the most general theory of linear perturbations around a given FRW evolution, characterised by a scale factor function $a(t)$. We will require that the FRW spacetime is geodesically complete for gravitons (other species may couple to a different effective metric) in the past and in the future. We will see (Section \ref{stability}) that one can identify a general gradient instability which will switch on somewhere along the history, unless a particular operator in the EFT, namely $\upleft{3}{R} \delta N$, is turned on, with a coefficient which changes sign along the way. This operator is typical of the so-called beyond-Horndeski theories \cite{Gleyzes:2014dya} (see \cite{Zumalacarregui:2013pma} for an earlier study). Usually one can dispose of the operator $\upleft{3}{R} \delta N$ when the coupling with matter is of no interest: it is a redundant operator that can be removed by a generalised disformal transformation \cite{Gleyzes:2014qga}. This transformation, however, becomes singular when the coefficient changes sign (see Appendix \ref{appdisformal}), so that for our purposes there is no way to get rid of this operator which remains crucial to ensure stability. Notice that the operator $\upleft{3}R\delta N$ gives rise (for instance in Newtonian gauge \cite{Gleyzes:2013ooa}) to equations of motion with three derivatives. Even if this fact does not give rise to extra degrees of freedom or pathologies in the particular case at hand, it is impossible to have the $\upleft{3}R\delta N$ operator starting from a theory with only second-order equations of motion.

\vskip 0.15cm

Since we know that the universe contains many different species, it is mandatory to study what happens when they are considered. The simplicity of the EFT approach allows to generalise the study including an arbitrary number of fields and fluids (Section \ref{multifield}). The conclusions apply completely unchanged to the adiabatic mode of the whole system. Some examples of stable FRW evolution are given in Section \ref{examples}, while Section \ref{outlook} is devoted to the conclusions.

\vskip 0.15cm

{\bf Note added:} While this paper was close to conclusion, the preprint \cite{Cai:2016thi} appeared with substantial overlap with our work.

\section{EFT for Single-Field Non-Singular Universe}
\label{formalism}
Consider a flat FRW universe that evolves throughout its history, spanning the cosmic time interval $-\infty \leq t \leq \infty$, without ever developing a singularity (by `singularity' we mean divergence of any physical quantity, such as e.g.~the expansion rate). We will further assume in this Section that the cosmological evolution is driven by a single scalar $\phi(t,\vec x)$ that acquires a time-dependent expectation value $\bar\phi(t)$. This configuration spontaneously breaks the time translation-invariance of the underlying theory, but leaves the (time-dependent) spatial diffeomorphisms, $x^i\to x^i+\xi^i(t,\vec x)$, intact. In the \textit{unitary gauge} defined by frozen scalar fluctuations $\phi(t,\vec x)=\bar\phi(t)$, the most general action consistent with this symmetry breaking pattern is \cite{Creminelli:2006xe}
\begin{equation}
\label{EFTI-action-Eij}
\begin{split}
S &=  \int\D^4x \, N\sqrt{h} \bigg[
	\frac{1}{2}M_{\rm Pl}^2\left(\upleft{3}{R} + \frac{E_{ij}E^{ij} - E^2}{N^2} \right)
	- \frac{M_{\rm Pl}^2\dot{H}}{N^2} - M_{\rm Pl}^2(3H^2+\dot{H}) 
\\
&+ \frac{1}{2}m_2^4\delta N^2 
	- \hat{m}_1^3\delta N\delta E
- \frac{1}{2}\bar{m}_1^2\delta E^2
	- \frac{1}{2}\bar{m}_2^2\delta {E^i}_j\delta {E^j}_i 
	-\bar m_3^2 \upleft{3}{R} \delta N+ \ldots
\bigg] \, .
\end{split}
\end{equation}
In writing this action we have made use of the ADM decomposition of the metric 
\be
\rd s^2 = -N^2 \rd t^2+h_{ij}(\rd x^i+N^i \rd t)(\rd x^j+N^j \rd t)~,
\ee
assuming a background solution of the FRW form: $\bar N = 1$, $\bar N^i=0$, $\bar h_{ij}=a^2(t)\delta_{ij}$.
Furthermore, $E_{ij}$ is related to the extrinsic curvature of the uniform-time hypersurfaces as
\be
E_{ij}\equiv N K_{ij}= \frac{1}{2}\(\dot h_{ij}-\nabla_i N_j-\nabla_j N_i\)~,
\ee
where $\nabla_i$ is the covariant derivative associated with the spatial metric $h_{ij}$, and by $\delta$ we denote the perturbation of the relevant quantity over its background value (e.g., $\delta N = N-1$, $\delta E = E-3H$, etc.). By the ellipses in \eqref{EFTI-action-Eij} we  denote terms of higher order in field perturbations and/or spacetime derivatives.
Notice that only the operators on the first line of \eqref{EFTI-action-Eij} feature tadpoles (that is, terms linear in perturbations), whose cancellation unambiguously determines the coefficients of these operators. All the rest of the operators are manifestly at least quadratic in perturbations, and their coefficients are free \cite{Cheung:2007st}. Furthermore, the symmetry breaking pattern at hand allows these to be functions of time.  In an inflationary quasi-de Sitter universe, one usually considers the EFT coefficients to be weakly time-dependent (e.g., $\rd m_2^4/\rd t \ll H m_2^4$); however, in what follows we will also be interested in regimes far from de Sitter, where these coefficients acquire strong time-dependence. Despite appearance, all of the EFT coefficients can have either sign (e.g. $m_2^4$ can be both positive and negative): our notation only serves to highlight their mass dimension. 

\vskip 0.15cm

As we will see, the unitary-gauge action \eqref{EFTI-action-Eij} leads to equations of motion which are second-order in time for the propagating fields -- the scalar and the graviton (see e.g. \cite{Gleyzes:2013ooa}). Therefore, no issues related to Ostrogradsky ghosts can arise.\footnote{Recently, new theories which avoid Ostrogradski instabilities have been proposed \cite{Langlois:2015cwa,Langlois:2015skt}, which may or may not be captured by the action \eqref{EFTI-action-Eij}.}  For a particular choice of the parameters
\be
\label{galileons}
\bar m^2_1=-\bar m^2_2~, 
\ee
these equations are also second-order in space. In this paper, we will be interested in precisely the latter case. Non-singular cosmologies based on theories that do not satisfy \eqref{galileons} (such as the \textit{ghost condensate} \cite{ArkaniHamed:2003uy}) have been studied in Ref. \cite{Creminelli:2007aq}, and we refer the reader to this reference for more detail. 

\vskip 0.15cm

For $\bar m^2_i$ vanishing, \eqref{EFTI-action-Eij} captures the dynamics of the quadratic perturbations in the (generalized cubic Galileon) theories studied in the context of non-singular cosmologies in Refs. \cite{Pirtskhalava:2014esa,Libanov:2016kfc}. The specific case of non-zero coefficients $\bar m_i^2$, satisfying
\be
\label{horndcond}
\bar m^2_2=-\bar m^2_1=2\bar m_3^2
\ee
 corresponds to the more general Horndeski/generalized Galileon models \cite{Nicolis:2008in,Horndeski:1974wa}, while the so-called theories \textit{beyond Horndeski} \cite{Gleyzes:2014dya} are captured by the same action \eqref{EFTI-action-Eij}, but with $\bar m_3^2$ unrelated to $\bar m_1^2  = -\bar m_2^2$. In the latter case, the equations of motion generically become third-order away from the unitary gauge, nevertheless the Ostrogradski ghost is absent \cite{Gleyzes:2013ooa}. 

\vskip 0.15cm

It is important to note that we have defined our action \eqref{EFTI-action-Eij} in the frame where the coefficient of the first operator---the 4D Einstein-Hilbert term---is the usual time-independent Planck mass. In contrast, non-minimally coupled theories such as Horndeski/generalized Galileons generically lead to a \textit{time-dependent} effective Planck scale on cosmological backgrounds. Putting them in the form \eqref{EFTI-action-Eij} requires an additional conformal transformation on the metric (followed by a reparametrization of time to bring back the expectation value of the lapse function to one) -- see Appendix \ref{appdisformal} for more detail. For this transformation to be non-singular, it is important that the effective Planck mass \textit{before} transition to the frame \eqref{EFTI-action-Eij} be a strictly positive function of time everywhere on the temporal domain. Violation of this condition would imply vanishing of the tensor modes' gradient energy at some point in time. Conversely, for solutions with a strictly positive effective Planck mass in the original frame, the tensor modes are manifestly stable and weakly coupled and the transformation to the frame \eqref{EFTI-action-Eij} is well-defined.\footnote{This explains why our analysis does \textit{not} capture the recently proposed non-singular scenarios that feature tensor modes with asymptotically vanishing kinetic terms \cite{Ijjas:2016vtq}.}  

\vskip 0.15cm

The effective theory \eqref{EFTI-action-Eij} can be further simplified by means of a perturbative redefinition of the fields. Indeed, for theories satisfying \eqref{galileons} it is possible to set 
\be
\label{zero}
\bar m_1^2  = -\bar m_2^2 = 0
\ee
with the help of a conformal + disformal transformation of the metric, described in Appendix \ref{appdisformal}. This amounts to moving into the frame where the graviton propagates at unit speed \cite{Creminelli:2014wna}. We will refer to the latter as the \textit{Einstein frame} and will exclusively work in this frame in what follows. The virtue of this frame is that the propagation of the graviton is completely standard, and one can easily check geodesic completeness. 

\vskip 0.15cm

Within the Einstein frame, it is generically possible to perform another, generalized disformal transformation that would make also the coefficient $\bar m_3^2$ vanish, see Appendix \ref{appdisformal}. (This is the linearized version of the transformation that is used to establish equivalence of some of the theories beyond Horndeski and Horndeski/generalized Galileons \cite{Gleyzes:2014qga}.) However, crucially for the discussion to come, in certain physically relevant cases the latter transformation becomes singular and can not be implemented.

\subsection*{The spectrum}
The dynamical degrees of freedom are conveniently described in the $\zeta$ \textit{gauge} \cite{Maldacena:2002vr}:
\be
h_{ij}=a^2 e^{2\zeta} \(e^{\gamma}\)_{ij}~, \qquad \gamma_{ii} = \p_i\gamma_{ij}=0~,
\ee
where $\zeta$ and $\gamma$ capture the dynamics of the scalar and the tensor degrees of freedom respectively. 
The action for $\zeta$---the curvature perturbation on comoving hypersurfaces---can be obtained by integrating out the non-dynamical lapse ($N$) and shift ($N^i$) variables from the Hamiltonian and momentum constraint equations in the standard way. For vanishing $\bar m_3^2$, this has been done in \cite{Creminelli:2006xe,Pirtskhalava:2014esa} and here we generalize the results of these references to the case of a non-zero $\bar m_3^2$, assuming we are in the Einstein frame \eqref{zero}. A straightforward calculation yields 
\begin{equation}
\label{zetaaction}
S_\zeta=\int \D^4x~ a^3\bigg[A\dot\zeta^2-B\left(a^{-1}\vec{\nabla}\zeta\right)^2 \bigg]~,
\end{equation}
where the two (generically time-dependent) coefficients read
\begin{align}
\label{A'}
A &=M_{\rm Pl}^2 \cdot \frac{3(2M_{\rm Pl}^2 H-\hat m_1^3)^2+ 2 M_{\rm Pl}^2 (m_2^4-2M_{\rm Pl}^2 \dot H-6M_{\rm Pl}^2 H^2) }{(2M_{\rm Pl}^2 H-\hat m_1^3)^2 }\\
\label{B'}
B&=-M_{\rm Pl}^2+\frac{1}{a} \cdot \partial_t Y ~,
\end{align}
where
\ba
Y\equiv a\cdot \frac{2 M_{\rm Pl}^2 (M_{\rm Pl}^2-2\bar m_3^2)}{2M_{\rm Pl}^2 H-\hat m_1^3} ~.
\label{Y}
\ea
This Lagrangian thus propagates a single, unitary scalar degree of freedom, provided that the quantities $A$ and $B$ are positive \cite{Gleyzes:2013ooa}.
The dynamics of tensor perturbations are exacty those of General Relativity by our choice of the frame \eqref{zero}.

\section{Stability}
\label{stability}

One can easily arrange for the kinetic coefficient $A$ in \eqref{A'} to be positive via imposing the following condition (which should of course hold at any particular moment of time)
\be
\label{posA}
3(2M_{\rm Pl}^2 H-\hat m_1^3)^2+ 2 M_{\rm Pl}^2 (m_2^4-2M_{\rm Pl}^2 \dot H-6M_{\rm Pl}^2 H^2)>0~.
\ee 
What remains is to make sure that also $B$ is positive, that is, the scalar perturbations are free from gradient instabilities throughout the entire evolution. It will prove convenient to put this condition, using \eqref{B'}, in the integrated form 
\be
\label{yeq}
Y(t_f)-Y(t_i) > \mpl^2 \int_{t_i}^{t_f} \rd t~ a(t)~,
\ee
which shows that $Y$ is a monotonically growing function of time. 
\vskip 0.15cm
Consider now a universe that at no point during its evolution encounters a singularity, that is, no physical quantity such as the Hubble rate or its derivatives ever diverges, while the scale factor may tend to zero or infinity asymptotically in time. Setting $t_{i}~(t_{f})\to -\infty ~(+\infty)$, the integral on the right hand side of \eqref{yeq} may or may not be finite for a finite $t_{f}~(t_{i})$. 
We are particularly interested in cosmologies for which the integral in \eqref{yeq} diverges on both ends. The condition 
\be
\int_{-\infty }^t dt ~ a(t) = \infty
\ee
is precisely the one expressing past-completeness of a given FRW cosmology \cite{Borde:2001nh,Libanov:2016kfc}, while future-complete cosmologies satisfy a similar condition, but with integration from a finite time to $t=+\infty$. Notice that this is only true in the Einstein frame where $\mpl = \rm{const}$ and the speed of graviton propagation is equal to one, in which case the graviton's affine parameter is related to the scale factor as $d\lambda = a(t) dt$ (simply a consequence of the gravitational redshift).
Therefore we are concentrating on spacetimes which are geodesically complete for the propagation of the gravitons.\footnote{The scalar perturbation $\zeta$ and other particles will in general move on a different effective metric, so that one should check geodesic completeness separately for each species. For example, in the case of $\zeta$, the relevant integral to look at is $\int dt~ a(t) B(t)$. One can check that this integral is invariant under both conformal and disformal transformations.}

\vskip 0.15cm

One example of a geodesically complete cosmology is a smooth bouncing universe, which starts out contracting from an asymptotically flat ($a(t\to-\infty)=\text{const}$) state; see e.g.~\cite{Ijjas:2016vtq} for a recent discussion. Another example is a universe that starts expanding from a Minkowski spacetime like in \textit{Galilean Genesis} \cite{Creminelli:2008wc} and smoothly transits into an inflationary de Sitter regime, where $a(t\to\infty)\propto e^{H_It}$. This transition has been studied in \cite{Pirtskhalava:2014esa} within a sub-class of \eqref{EFTI-action-Eij} (only $m_2^4$ and $\hat m_1^3$ are non-vanishing) and has been found to suffer from a gradient instability right before the onset of the inflationary phase.\footnote{Notice that backgrounds describing the transition between two asymptotic de Sitter spaces are not geodesically complete in the past. As discussed in Refs.~\cite{Pirtskhalava:2014esa} and \cite{Alberte:2016izw}, there are no generic issues with stability in this case (see also \cite{Libanov:2016kfc} for a different example).} The fact that the gradient instability is unavoidable for geodesically complete cosmological backgrounds in this class of theories has been recently formulated as a theorem in \cite{Libanov:2016kfc} and subsequently generalized to the broader class of Horndeski theories in \cite{Kobayashi:2016xpl}. In what follows, we will re-derive these no-go results\footnote{In comparing with the previous references notice that our formulas will be simpler since we chose to work in the Einstein frame from the beginning. As we discussed, this is always possible without loss of generality.} and will point out ways to evade them.

\vskip 0.15cm

Since the integral in \eqref{yeq}  diverges on both ends $Y$ has to start off at $Y(t\to -\infty)= -\infty$ and go to $Y(t\to \infty)= \infty$, so that it has to cross zero at some $t=t_0$. In 
Horndeski theories $\bar m_3^2=0$ and the expression in \eqref{Y} further simplifies to
\be
\label{horndy}
Y = 2 a \cdot \frac{\mpl^4}{2\mpl^2 H -\hat m_1^3}~.
\ee
It thus becomes evident that, assuming a continuous $Y$, the gradient instability in the scalar sector can only be avoided if the denominator of \eqref{horndy} diverges at $t=t_0$, implying a singularity \cite{Libanov:2016kfc,Kobayashi:2016xpl}. Alternatively, one can consider the possibility of a discontinuous $Y$, which corresponds to the denominator in \eqref{horndy} vanishing at one point. However, one can check that the coefficients $A$ and $B$ of the $\zeta$ action would diverge in this case. 

\vskip 0.15cm

The above discussion also suggests that the no-go result can be evaded if the parameter $\bar m_3^2$ does not vanish, as it happens e.g.~in theories beyond Horndeski. Indeed,  in this case,  as it follows from \eqref{Y}, $Y$ can continuously pass through zero without any pathology if the following relation becomes true at some time
\be
\label{ycond}
 \bar m_3^2=\frac{\mpl^2}{2}~.
\ee
This establishes that the conclusions of the theorems of \cite{Libanov:2016kfc,Kobayashi:2016xpl} can be avoided by a slight generalization of the domain of theories under consideration.

\vskip 0.15cm

We have seen that stable geodesically complete cosmologies require a non-zero coefficient $\bar m_3^2$ for the operator $ \upleft{3}{R} \delta N$.
However usually this operator can be removed by a suitable generalised disformal transformation \cite{D'Amico:2016ltd}. One could thus wonder how this can be reconciled with the inevitable presence of instabilities in theories with $\bar m_3^2=0$.
This apparent paradox has a simple resolution. The transformation required to get rid of $\bar m_3^2$ becomes singular precisely when $\bar m_3^2$ crosses zero and this, as we discussed, is necessary for the stability of the solution, see the Appendix \ref{appdisformal} for a more detailed discussion on these matters. Therefore stable geodesically complete cosmologies require the operator $ \upleft{3}{R} \delta N$. This operator, depending on the gauge, gives rise to equations of motion with more than two derivatives. Therefore it cannot appear starting from standard theories with second order equations of motion.

\vskip 0.15cm

In Sec.~\ref{examples} we provide two explicit examples of smooth geodesically complete cosmologies which illustrate how the theorems of Refs. \cite{Libanov:2016kfc} and \cite{Kobayashi:2016xpl} can be evaded.

\section{Generalization to Multiple Fields}
\label{multifield}

One of the main advantages of the EFT method is that it allows an immediate generalization of our conclusions to the multifield case. (The EFT of perturbations in the presence of more than one field has been studied in the context of inflation in \cite{Senatore:2010wk}.) The bottom line is very simple: the conclusions above remain unaltered even when we have a system which includes many fields and energy components. (Reaching the same conclusion starting from a putative multifield generalization of beyond-Horndeski theories coupled to other fluids looks prohibitively difficult. A particular two-field case was studied in \cite{Kolevatov:2016ppi}, finding that the second field could not fix the instability.) 
In the presence of many fields (or fluids) the EFT we described in the previous Sections still captures the dynamics of the adiabatic mode, the one in which all the components fluctuate in unison, locally reproducing the unperturbed FRW evolution. Additional fields will couple to the EFT of eq.~\eqref{EFTI-action-Eij} and may have kinetic mixings with the adiabatic mode. However, the existence, under the assumptions spelled out before, of a gradient instability cannot be cured by the existence of these additional fields: the instability in the adiabatic direction remains and can be analyzed setting all the additional fields to zero in the first place. 

\vskip 0.15cm

Let us make an example to clarify the general argument. Consider a field, named $\sigma$, which describes an additional perturbation on top of the adiabatic one. It will couple to the theory \eqref{EFTI-action-Eij} in a way compatible with the symmetries. For simplicity we assume a shift symmetry $\sigma \to \sigma + c$. Since $\sigma =0$ must be a solution, one is not allowed to write any tadpole term for it, while, at lowest order in derivatives, the only mixing term is of the form $\delta N \dot\sigma$. On top of that one has terms quadratic in $\sigma$: for simplicity let us take $\sigma$ to have a Lorentz invariant kinetic term, although generalizations are straightforward. The new terms in the action are
\begin{equation}
S_\sigma = \int\D^4x \, N\sqrt{h} \bigg[
		m_\sigma^2(t)\delta N \dot{\sigma} -\frac12 g^{\mu\nu}\partial_\mu\sigma \partial_\nu\sigma + \ldots \bigg] ~.
\label{EFT-multi-fields}
\end{equation}
The solution of the constraints can be carried out as before and it gives
\begin{equation}
S_\zeta=\int \D^4x~ a^3\bigg[A(t)\dot\zeta^2-B(t)\left(a^{-1}\vec{\nabla}\zeta\right)^2 + E(t)\dot{\sigma}\dot{\zeta} + \frac12 \dot\sigma^2 - \frac{1}{2} (a^{-1}\vec \nabla\sigma)^2\bigg] ~.
\end{equation}
The expression for $A$ and $B$ remains the same as in the absence of $\sigma$, eq.s \eqref{A'} and \eqref{B'}, while $E$ is given by
\begin{equation}
\label{DEFm}
E=\frac{2 M_{\rm Pl}^2 m_\sigma^2 }{2M_{\rm Pl}^2 H-\hat m_1^3 } ~.
\end{equation}
It is clear that the new time kinetic mixing term does not change in any way the conclusions reached in the previous Sections\footnote{As pointed out in \cite{Senatore:2010wk}, the symmetries of the EFT do not allow to write a spatial kinetic mixing between $\sigma$ and $\zeta$. However, even if it were possible, it would not change our conclusions, since the mixing cannot avoid the instability in the $\zeta$ direction.}.  Notice that this formulation is very general and for example it applies to the original Genesis scenario \cite{Creminelli:2008wc} which features a singularity $H \to +\infty$. Before reaching the singularity the EFT breaks down and one can imagine a transfer of energy to a thermal bath of particles, i.e.~reheating. However, once we take into account all the components, the same conclusions as discussed above can be reached about the overall adiabatic mode.

\section{Examples}
\label{examples}
 In this Section we provide several explicit examples of stable, geodesically-complete cosmologies described in Sec.~\ref{stability}. We will show how the gradient instability discussed in Refs. \cite{Libanov:2016kfc} and \cite{Kobayashi:2016xpl} can be avoided in effective theories of the kind \eqref{EFTI-action-Eij} for which the condition \eqref{horndcond} (which in the Einstein frame simply implies a vanishing $\bar m_3^2$) fails to hold, as it happens e.g. in theories beyond Horndeski. Furthermore, we will confirm our expectations based on the discussion of Sec.~\ref{stability} whereby stability of the system requires that the (time-dependent) coefficient $\bar m_3^2$ satisy Eq.~\eqref{ycond} at a certain point in time. This implies that the given backgrounds \textit{cannot} be smoothly deformed into a frame where the coefficient of the would-be redundant operator $\upleft{3}{R} \delta N$ vanishes. 

\subsection{A smooth bounce}

\vskip 0.15cm
Our first example is a bounce recently discussed in Ref. \cite{Ijjas:2016tpn} and defined by the following profile for the Hubble rate
\be
\label{bounce}
H(t) = H^2_0 t e^{-\alpha (t-t_\star)^2}~,
\ee
where $H_0,~\alpha$ and $t_\star$ are constant parameters. The ansatz \eqref{bounce} describes a universe that starts slowly contracting from an asymptotically Minkowski state, undergoes a bounce at $t=0$ and expands, again approaching the Minkowski spacetime in the asymptotic future. Importantly, the Lagrangian of the model that realizes this cosmology in \cite{Ijjas:2016tpn} corresponds to $\bar m_3^2 =0$ in the language of the effective theory \eqref{EFTI-action-Eij},  and thus conforms to the assumptions of the no-go theorems of Refs. \cite{Libanov:2016kfc,Kobayashi:2016xpl}. Therefore, it has unsurprisingly been found to suffer from gradient instability \cite{Ijjas:2016tpn}.\footnote{The authors of \cite{Ijjas:2016tpn} argue that one can nevertheless keep the instability away from the bounce. Moreover, in a more recent work \cite{Ijjas:2016vtq} they have come up with an example of an extended theory that realizes bouncing cosmology in a manifestly stable way, albeit at the expense of asymptotically vanishing kinetic coefficients of the tensor perturbations. Notice that this model does not comply with our assumption of geodesic completeness.} Here we wish to show how a fully stable and subluminal realization of \eqref{bounce} arises for a non-zero $\bar m_3^2$. 

\begin{figure}
\centering
\includegraphics[width=.48\textwidth]{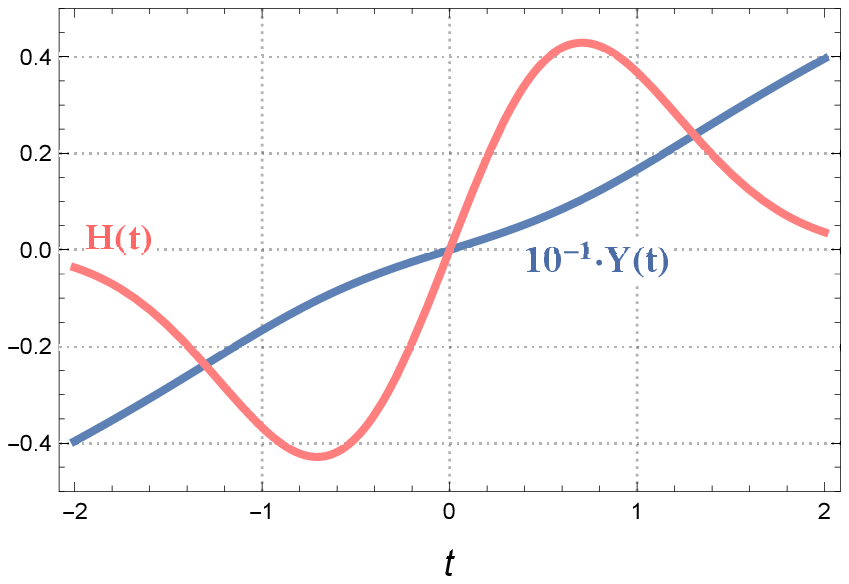} \quad
\includegraphics[width=.48\textwidth]{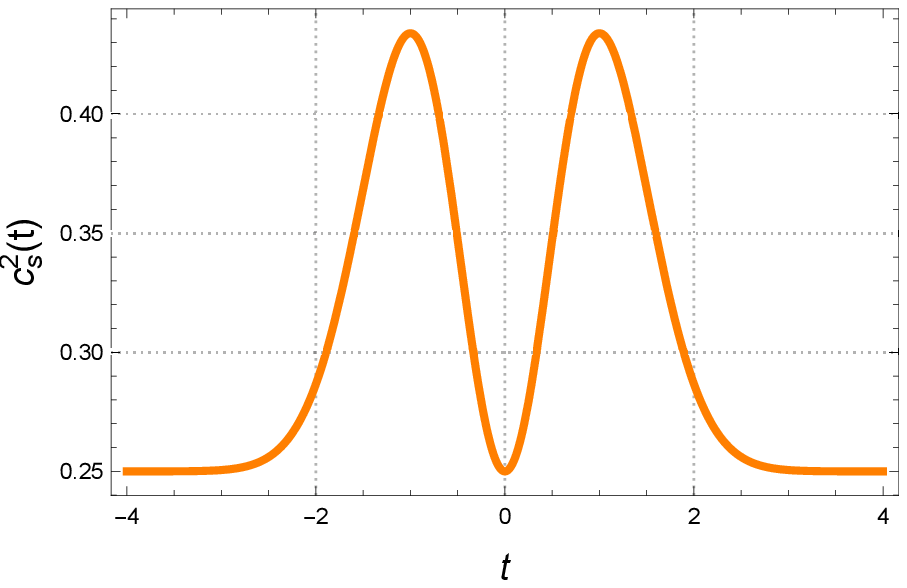}
\caption{An example of a fully stable and subluminal bounce, described by the profile \eqref{bounce} for the Hubble rate. The EFT coefficients have been chosen as in \eqref{bounceparameters}. The plots correspond to the following choice of the model parameters $(\mpl = 1)$: $H_0=1,~t_\star=0,~\alpha=1$, $\kappa=2,~\beta = 1/2-t^2 e^{-t^2}$. The kinetic coefficient $A$ in \eqref{zetaaction} is constant ($A=4$) for the given ansatz, while the time-dependent $\zeta$-speed of sound (whose asymptotic value is $c_s^2=1/4$) is shown on the right panel. The left panel illustrates the behavior of the function $Y$, which passes through zero at $t=0$, as required by stability of the system.}
\label{figbounce}
\end{figure}

\vskip 0.15cm

To this end, consider the following choice of the EFT coefficients in the Einstein frame ($\bar m_{1,2}^2 = 0$)
\be
\label{bounceparameters}
\hat m_1^3=0,\quad m_2^4=\kappa \mpl^2 H^2 + 2\mpl^2 \dot H+6 \mpl^2 H^2,\quad \bar m_3^2 = \beta \mpl^2~.
\ee
The corresponding theory is determined by two free dimensionless functions of time, $\kappa$ and $\beta$, which can always be chosen such that the scalar perturbation $\zeta$ is fully stable and (sub)luminal.  One such choice of the parameters is shown in Fig. \ref{figbounce}. The scalar perturbations propagate subluminally all along the bounce \eqref{bounce} in this model, with their speed of sound approaching $c_s^2\to 1/4$ in both asymptotics. 

\subsection{From Minkowski to de Sitter}

Another example of a geodesically-complete cosmology we wish to consider here is a Minkowski $\to$ de Sitter transition \cite{Pirtskhalava:2014esa}, whereby a NEC-violating universe initially expanding out of a zero-curvature state ends up in an inflationary quasi-de Sitter phase. As the corresponding backgrounds fall into the type I category according to the classification of Sec. \ref{stability}, they have been found to suffer from gradient instability \cite{Pirtskhalava:2014esa,Libanov:2016kfc,Kobayashi:2016xpl} within a rather broad spectrum of models, including Horndeski/generalized Galileons. We wish to show how the instability can be fixed in a slightly more general set of theories with a non-zero $\bar m_3^2$.
\begin{figure}[t!]
\center
\includegraphics[width=.48\textwidth]{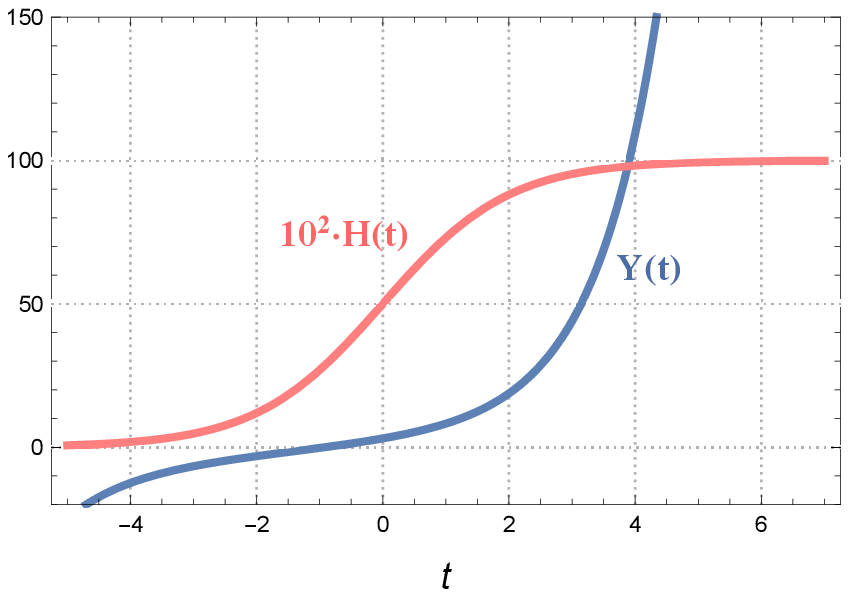} \quad
\includegraphics[width=.48\textwidth]{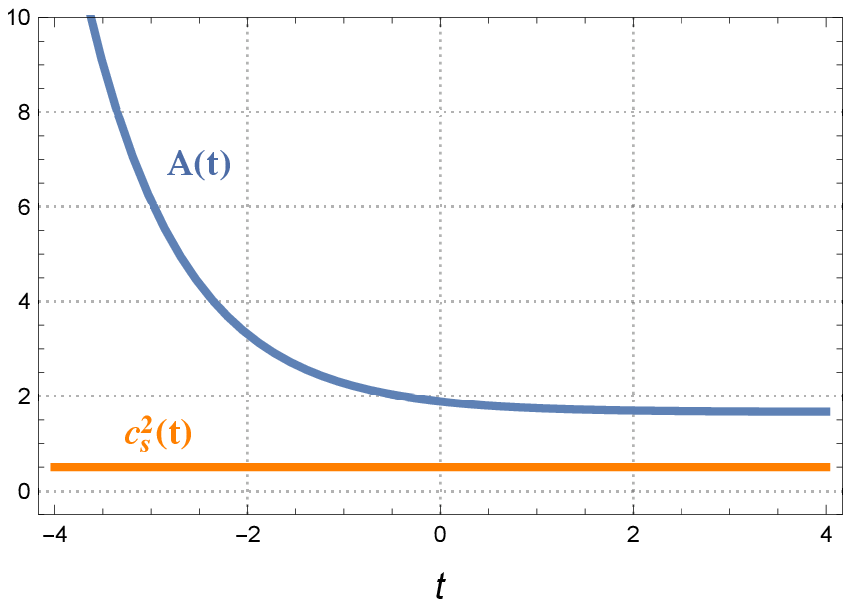}
\caption{An example of a fully stable and subluminal background that interpolates between Minkowski and de Sitter spacetimes, described by the profile \eqref{minkds} for the scale factor. The EFT coefficients have been chosen as in \eqref{minkdseftcoeffs}. The plots correspond to the following choice of the model parameters $(\mpl = 1)$: $H_I=1,~a_0=1,~a_1=1$, $\kappa=3,~\xi=-1$, $\beta = -(7/24)\cdot (5 t e^t +3 e^{2t}-2)/(e^t+1)^2$. The kinetic coefficient $A$ and the $\zeta$-speed of sound are shown on the right panel. Notice that $A$ diverges in the asymptotic past (like it does e.g.~in Galilean Genesis \cite{Creminelli:2010ba}), which makes the scalar perturbations  weakly coupled. Nonetheless, $c_s^2$ is always finite (for the ansatz at hand it is in fact constant at all times, $c_s^2=1/2$). The left panel illustrates the behavior of the function $Y$, which passes through zero at $t<0$, as required by the stability of the system.}
\label{figminkds}
\end{figure}
\vskip 0.15 cm 
To this end, consider the following ansatz for the scale factor 
\be
\label{minkds}
a = a_0+a_1 e^{H_I t}~,
\ee
where $a_{0}$ and $a_1$ are constant (and positive) parameters, while $H_I$ is the late-time expansion rate of the universe. We will also assume the following form of the Einstein-frame EFT coefficients 
\be
\label{minkdseftcoeffs}
 \hat m_1^3 = \xi \mpl^2 H, \quad m_2^4 = \kappa \mpl^2 \dot H, \quad \bar m_3^3 = \beta\mpl^2~,
\ee
so that the theories under consideration are determined by three free dimensionless functions of time: $\xi,~\kappa$ and $\beta$.

\vskip 0.15cm

The explicit examples of stable (and subluminal) backgrounds that interpolate between Minkowski and de Sitter spacetimes are shown in Fig.~\ref{figminkds}. Note that the speed of scalar perturbations for the given choice of the parameters is constant everywhere, $c_s^2=1/2$. Again, the ultimate reason for the complete stability of these theories lies in the effects of the operator $\upleft{3}R\delta N$.

\section{Outlook and Future Directions}
\label{outlook}
In this paper we have shown that fully stable spatially flat FRW solutions which are regular and geodesically complete are possible, albeit they require the existence of the somewhat exotic operator $\upleft{3}{R} \delta N$ in the EFT for perturbations with a coefficient which changes sign along the way. This operator is known to be asbent if one starts with a Horndeski theory -- the most general scalar-tensor theory with second-order equations of motion \cite{Gleyzes:2014dya,Gleyzes:2014qga}. However, $\upleft{3}{R} \delta N$ still does not seem to be associated to any pathology, at least from the EFT point of view and it arises in beyond-Horndeski theories, which are in general healthy despite having higher-order equations of motion \cite{Gleyzes:2014dya,Gleyzes:2014qga}. Another example of a system which can give rise to a completely smooth cosmology is the Ghost Condensate \cite{ArkaniHamed:2003uy,Creminelli:2006xe}. (This was not discussed in this paper since the equations of motion of the ghost condensate are of higher order in spatial derivatives.) One is left with a (somewhat vague) connection between these regular NEC-violating cosmologies and theories with higher derivatives but not propagating additional pathological degrees of freedom. It would be worthwhile looking into this connection in more detail, also taking into account the new examples which are even more general than beyond-Horndeski (see for example \cite{BenAchour:2016fzp}). Of course, the overarching problem is to understand the possibility of a standard UV completion in theories at hand: this problem is particularly severe since the stability arguments push us towards a rather exotic corner of the theory space.

\subsection*{Acknowledgements}
We would like to thank Filippo Vernizzi for useful discussions. The work of D.P. is supported by the Swiss National Science Foundation under grant 200020-150060. The work of E.T. is supported in part by MIUR-FIRB grant RBFR12H1MW.

\appendix

\section{The Various Frames and Their (In)equivalence}
\label{appdisformal}
In this appendix we elaborate on the field redefinition that takes us to the Einstein frame, corresponding to $\bar m_1^2=\bar m_2^2 = 0$ and characterized by the graviton propagating at unit speed. We will also comment on the possibility to further set $\bar m_3^2\neq 0$ by means of a disformal transformation. Our discussion closely parallels \cite{D'Amico:2016ltd} and we refer the reader to that reference (and references therein) for more details on the technicalities involved. 

\vskip 0.15cm
 
Our starting point is the EFT \eqref{EFTI-action-Eij}, explicitly expanded---including the first line---to the quadratic order in field perturbations (from now on we assume \eqref{galileons} holds)
\begin{equation}
\begin{split}
S^{(2)} = ~ & \frac{M^2_{\text{Pl}}}{2}\int\D^4x \, a^3 \bigg[
	 \left(-6H^2-2\dot{H}+\frac{m^4_2}{M^2_{\text{Pl}}}\right) \delta N^2 
	+\left(4H - \frac{2\hat{m}_1^3}{M^2_{\text{Pl}}}\right) \delta N\delta E
\\
&	- \left(1+\frac{\bar{m}_1^2}{M^2_{\text{Pl}}}\right) \(\delta E^2-\delta {E^i}_j\delta {E^j}_i \)
	+ \left(1-\frac{2\bar{m}_3^2}{M^2_{\text{Pl}}}\right) \upleft{3}{R} \delta N + \delta_2\left(\sqrt{h}\upleft{3}{R}/a^3\right)
\bigg] 
\\
\equiv & \int\D^4x \, a^3 \frac{M^2}{2}\bigg[\delta K^i_j\delta K^j_i-\delta K^2+
	 \alpha_\text{K}H^2 \delta N^2 
	+ 4\alpha_\text{B}H\delta N\delta K
	+ \left(1+\alpha_\text{H}\right) \upleft{3}{R} \delta N 
\\
&	+ \left(1+\alpha_\text{T}\right)\delta_2\left(\sqrt{h}\upleft{3}{R}/a^3\right)
\bigg] \, .
\label{app-EFT}
\end{split}
\end{equation}
Here $\delta_2$ refers to the second-order term in the perturbative expansion of the appropriate quantity.
To facilitate comparison with \cite{D'Amico:2016ltd}, we have rewritten the last line of \eqref{EFTI-action-Eij} in terms of the extrinsic curvature, $\delta K_{ij} = \delta E_{ij}-  h_{ij} H\delta N $, and have defined
\begin{align}
M^2(t) &\equiv  \mpl^2+\bar m_1^2~,\label{M}\\
\alpha_\text{K}(t)
&	\equiv \(-\frac{2\dot{H}}{H^2} + 
\frac{m^4_2}{M^2_{\text{Pl}}H^2}
- \frac{6\hat{m}_1^3}{M^2_{\text{Pl}}H}
- \frac{6\bar{m}_1^2}{M^2_{\text{Pl}}}
\)\(1+\frac{\bar m_1^2}{\mpl^2}\)^{-1}~,
\label{aK}\\
\alpha_\text{B}(t)
&	\equiv 
-\(\frac{\hat{m}_1^3}{2M^2_{\text{Pl}}H}
+\frac{\bar{m}_1^2}{M^2_{\text{Pl}}}
\)\(1+\frac{\bar m_1^2}{\mpl^2}\)^{-1}~,
\label{aB}\\
\alpha_\text{H}(t)
&	\equiv -1 +\frac{\mpl^2-2\bar{m}_3^2}{\mpl^2+\bar m_1^2}~,
\label{aH}\\
\alpha_\text{T}(t)
&	\equiv -1 + \frac{M^2_{\text{Pl}}}{M^2_{\text{Pl}}+\bar{m}_1^2} \, ~.
\label{aT}
\end{align}
It follows from Eqs.~\eqref{horndcond} and \eqref{aH} that Horndeski theories have $\alpha_\text{H}=0$, while theories beyond Horndeski are characterized by a non-zero $\alpha_\text{H}$. The statement about equivalence of the two theories reduces to the fact that in most cases (but not always, as we will see below) there is a field redefinition that sets a generic $\alpha_\text{H}$ to zero \cite{Gleyzes:2014dya}.

\vskip 0.15cm

We will be interested in the transformation properties of the various EFT coefficients under a conformal $+$ disformal transformation of the form
\begin{equation}
g_{\mu\nu} \rightarrow C(t)g_{\mu\nu}+D(t,N)\delta_\mu^0\delta_\nu^0 \, ~.
\label{cttapp}
\end{equation}
Note that \eqref{cttapp} generically changes the expectation value $\bar N$ of the lapse. We will always assume $\bar N =1$, so that \eqref{cttapp} will be understood to be followed by an appropriate reparametrization of time to enforce this condition. With these qualifications, the parameters \eqref{M} - \eqref{aH} transform as follows \cite{D'Amico:2016ltd}
\begin{align}
M^2&\rightarrow \frac{M^2}{C\sqrt{1+\alpha_\text{D}}}\, \label{M1}\\
\alpha_\text{K}
&	\rightarrow \frac{\alpha_\text{K} + 12 \alpha_\text{B}\alpha_\text{CDX} - 6\alpha_\text{CDX}^2}{(1+\alpha_\text{CDX})^2} \, ,
\label{aK1}\\
\alpha_\text{B}
&	\rightarrow   \frac{1+\alpha_\text{B}}{1+\alpha_\text{CDX}}- 1 \, ,
\label{aB1}\\
\alpha_\text{H}
&	\rightarrow \frac{\alpha_\text{H}-\alpha_\text{X}}{1+\alpha_\text{X}} \, ,
\label{aH1}\\
\alpha_\text{T}
&	\rightarrow (1+\alpha_\text{T})(1+\alpha_\text{D}) -1 \, ,
\label{aT1}
\end{align}
where we have defined
\begin{equation}
\alpha_\text{C} \equiv \frac{\dot{C}}{2HC} \, ,
~
\alpha_\text{D} \equiv \frac{D}{C-D} \, ,
~
\alpha_\text{X} \equiv -\frac{1}{2C}\frac{\partial D}{\partial N} \,, ~ \alpha_\text{CDX} \equiv (1+\alpha_\text{C})(1+\alpha_\text{D})(1+\alpha_\text{X}) -1~ . 
\label{alphas}
\end{equation}
All quantities in the above expressions are evaluated using the background expectation value for the metric.

\vskip 0.15cm

Suppose now that one starts with a generic EFT of the kind \eqref{EFTI-action-Eij} with $\bar m_1^2 = -\bar m_2^2 \neq 0$. One can show that the graviton propagates at a speed $c_T^2 = \(1+\bar m_1^2/\mpl^2\)^{-1} \neq 1$ in this frame. Now, we will assume that $\bar m_1^2 \neq -\mpl^2$, that is,  the tensor perturbations are not infinitely strongly coupled ($c_T\neq \infty$). Then, the expressions \eqref{M}, \eqref{aT}, \eqref{M1} and \eqref{aT1} guarantee that there will always exist a non-singular transformation with
\be
C = c_T^{-1}, \qquad \alpha_D = c_T^{-2}-1~
\ee
that brings one to the frame where the graviton propagates at a unit speed $c_T^2\big |_{\rm new}=1$. We have assumed this frame everywhere above. 

\vskip 0.15cm

In this new frame, the coefficient $\bar m_3^2 = -\mpl^2\alpha_H/2$ (see Eq. \eqref{aH}) may or may not be equal to zero. Even if it does not vanish, it is generically possible to find a disformal transformation with $\alpha_{\rm X}\neq 0$ in \eqref{alphas} that sets $\bar m_3^2$ (and therefore $\alpha_H$) to zero in the transformed frame. This in particular amounts to mapping a beyond-Horndeski theory to Horndeski/generalized Galileons, as discussed above. 
It follows from \eqref{aH1}, that the transformation that apparently does the job has $\alpha_\text{X}(t)=\alpha_{\rm H}(t)\big |_{\rm old~frame}$ for all $t$.
However, we have argued above that stable non-singular type I cosmologies require $\bar m_3^2 = \mpl^2/2$, and therefore $\alpha_{\rm X}=-1$ to become true at least at one point on the temporal domain. Having $\alpha_{\rm H}$ vanish in the new frame would then imply that the numerator of \eqref{aH1} is a higher-order zero than the denominator at this point. Even if this is true, however,  $\alpha_{\rm X}=-1$ yields $\alpha_{\rm CDX}=-1$ in \eqref{alphas}, so that the coefficients $\alpha_{\rm K}$ and $\alpha_{\rm B}$ in the new frame inevitably blow up at the point under consideration, as follows from Eqs. \eqref{aK1} and \eqref{aB1}.
This shows that the would-be transformation that removes the $\upleft{3}R\delta N$ operator from the effective theory \eqref{EFTI-action-Eij} may become ill-defined in physically relevant situations, of which the geodesically-complete backgrounds discussed in Secs. \ref{stability} and \ref{examples} are an example.

\renewcommand{\em}{}
\bibliographystyle{utphys}
\addcontentsline{toc}{section}{References}
\bibliography{eftinf}

\end{document}